\documentclass[conference]{IEEEtran}
\IEEEoverridecommandlockouts
\usepackage{cite}
\usepackage{amsmath,amssymb,amsfonts}
\usepackage{algorithmic}
\usepackage{graphicx}
\usepackage{textcomp}
\usepackage{xcolor}
\def\BibTeX{{\rm B\kern-.05em{\sc i\kern-.025em b}\kern-.08em
    T\kern-.1667em\lower.7ex\hbox{E}\kern-.125emX}}

\usepackage{flushend}
\usepackage{subfig}
\usepackage{siunitx}
\renewcommand{\vec}[1]{\mathbf{#1}}

\begin{document}
\title{Data-Driven Models for Control Engineering Applications Using the Koopman Operator}

\author{\IEEEauthorblockN{Annika Junker\IEEEauthorrefmark{1}\thanks{\IEEEauthorrefmark{1}Corresponding author}, 
		Julia Timmermann\IEEEauthorrefmark{2} and
		Ansgar Trächtler\IEEEauthorrefmark{3}}
\IEEEauthorblockA{\textit{Heinz Nixdorf Institute,
	Paderborn University, Germany}\\
Email: \IEEEauthorrefmark{1}annika.junker@hni.upb.de,
\IEEEauthorrefmark{2}julia.timmermann@hni.upb.de,
\IEEEauthorrefmark{3}ansgar.traechtler@hni.upb.de}
%\and
%\IEEEauthorblockN{Julia Timmermann}
%\IEEEauthorblockA{\textit{Heinz Nixdorf Institute} \\
%\textit{Paderborn University}\\
%Paderborn, Germany \\
%julia.timmermann@hni.upb.de}
%\and
%\IEEEauthorblockN{Ansgar Trächtler}
%\IEEEauthorblockA{\textit{Heinz Nixdorf Institute} \\
%\textit{Paderborn University}\\
%Paderborn, Germany \\
%ansgar.traechtler@hni.upb.de}
}

\maketitle

\begin{abstract}
Within this work, we investigate how data-driven numerical approximation methods of the Koopman operator can be used in practical control engineering applications. We refer to the method Extended Dynamic Mode Decomposition (EDMD), which approximates a nonlinear dynamical system as a linear model. This makes the method ideal for control engineering applications, because a linear system description is often assumed for this purpose. Using academic examples, we simulatively analyze the prediction performance of the learned EDMD models and show how relevant system properties like stability, controllability, and observability are reflected by the EDMD model, which is a critical requirement for a successful control design process. Subsequently, we present our experimental results on a mechatronic test bench and evaluate the applicability to the control engineering design process. As a result, the investigated methods are suitable as a low-effort alternative for the design steps of model building and adaptation in the classical model-based controller design method.
\end{abstract}


\section{Introduction}\label{sec:introduction}
For several years, machine learning has been widely used to enhance intelligence in technical systems. Within the field of control engineering, there are, apart from that, classical best practices based on physical models. Therefore, an innovative approach is to combine the physical models with data-driven methods in a sensible way. 

The Koopman operator enables an operator-theoretical view of dynamical systems. The dynamics of an autonomous nonlinear dynamical system can be described linearly by means of the linear but infinite dimensional Koopman operator. The resulting globally linear description of a nonlinear system is extremely powerful because it opens new possibilities to apply the fully elaborated system theory and design procedures of linear control engineering to nonlinear systems \cite{MB04}. In addition, linear systems offer significantly reduced computational effort for prediction, which is highly beneficial for an internal model in the controller or observer. Due to the fact that the Koopman operator is infinite dimensional, it is common for algorithmic applications to utilize a numerical approximation of the Koopman operator using data from measurement or simulation.

\begin{figure*}[!t]
	\centering
	\includegraphics[scale=0.9]{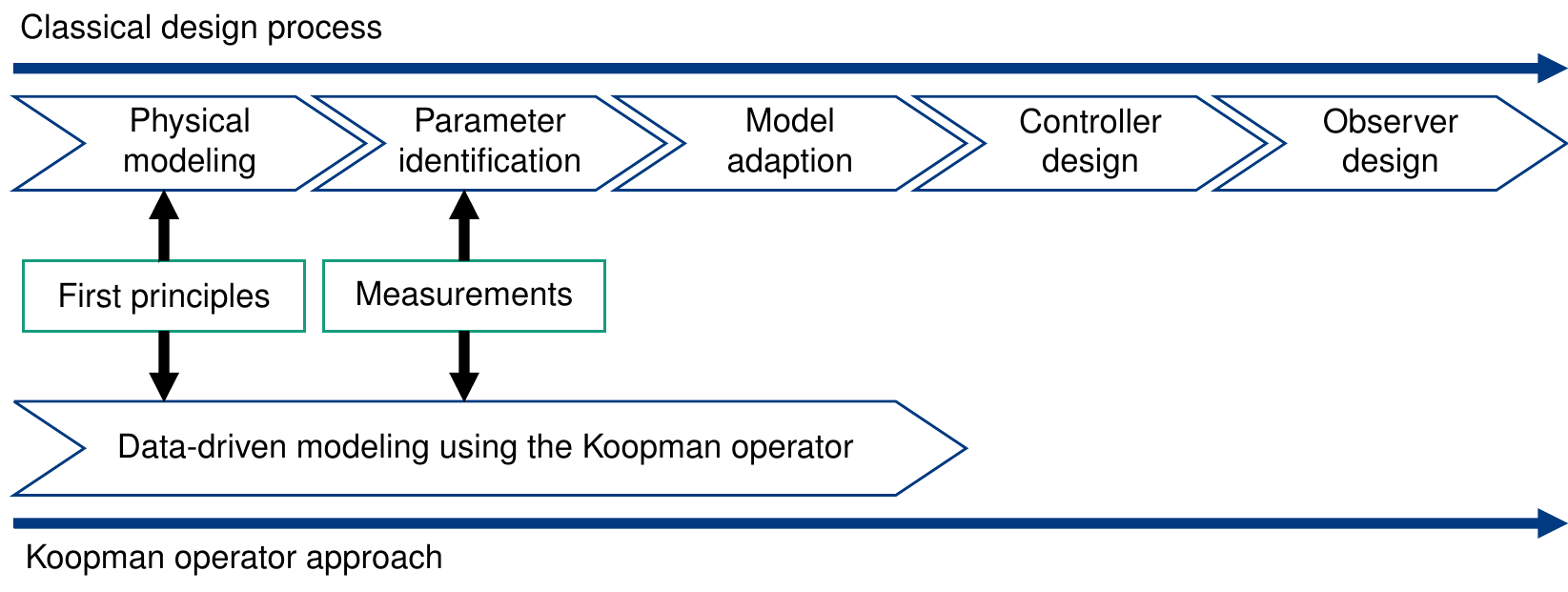}
	\caption{Koopman operator approach as an alternative to the classical model-based control engineering design process.}
	\label{fig:overview}
\end{figure*}

As shown in Fig. \ref{fig:overview}, Koopman operator-based models can be integrated into the classical model-based control engineering design process as an alternative approach. The classical design process consists of physical modeling based on first principles and parameter identification using measurement data from the real system, resulting in an accurate but typically nonlinear model. The subsequent controller and observer design often relies on linear control theory, so the nonlinear model must be linearized. In most cases, these steps are a big effort, so Koopman operator-based modeling can be considered as an alternative way as it directly provides a closed linear model, which can be used straight away for the controller and observer design. To successfully manage the application to control engineering systems, it is important that the data-driven models meet the following requirements:
\begin{itemize}
	\item high prediction accuracy with low computational effort
	\item correct representation of relevant system properties such as stability or characteristics of the eigenvalues, controllability and observability.
\end{itemize}
Furthermore, in the context of hybrid models, it is also advantageous if existing prior knowledge can be used in a sensible way. 

In recent years, several numerical techniques for the approximation of the Koopman operator have been developed. Most commonly, so-called snapshot methods are used, which extract dynamic information from an underlying system using a time-resolved sequence from measurement data. The most basic of these algorithms is Dynamic Mode Decomposition (DMD), which is closely related with Proper Orthogonal Decomposition (POD) and was first used in the flow fields community \cite{SMP09, PS09, Sch10, TRL+14}. A linear dynamical model is learned for the evolution of the original states of the system, thus the dynamic properties of this approach are very limited. The Extended Dynamic Mode Decomposition (EDMD) was developed to represent systems with strong nonlinearities. This method extends DMD by \textit{lifting} the original states of the system to a higher dimensional space using additional nonlinear observable functions. By learning the evolution of the \textit{lifted} states, this method yields a more precise approximation of the infinite-dimensional Koopman operator \cite{WKR15}. Also, control engineering systems have a rather small number of states, but with nonlinear dynamics.
Therefore, a higher dimensional space is readily acceptable if it leads to linear dynamics.  %Therefore, one would happily accept a higher dimensional space if it leads to linear dynamics.
Also related to the methods mentioned is SINDy (Sparse Idenfitication of Nonlinear Dynamics) \cite{BPK16}, which provides a strategy to extract the governing equations of a nonlinear dynamical system from data. Other approaches aim to determine a linear system description using neural networks \cite{AELM20, BK20, LKB18, LM20, OR17, RRW20, TKY17, vFKB20, YKH19, ZCKW18}. For application to controlled systems, the numerical approximation methods provide suitable extensions to learn the influence that the input has on the dynamics (DMDc, EDMDc, SINDYc, etc.) \cite{BPK16b, PBK16, WHD+16, PBK18}. In an alternative strategy, the control system is replaced by a finite set of autonomous systems by quantizing the control set \cite{PB21}.

There are applications to control systems already available underlying that the numerical approximation methods gain a high prediction accuracy and have the potential to be used for control design \cite{ALM17,CHM+20,KCH+20,MCTM19,MCTM20,ZB21}. Our approach extends these solutions by embedding the methods in the control engineering context and thus making them accessible to a broad application domain. For this purpose, we focus on ensuring relevant system theoretic properties in the data-driven model. 

The authors in \cite{MAM20} motivate the use of stability properties of the EDMD model to ensure approximation accuracy over a longer time horizon and for a small number of measured data by projecting the numerically computed Koopman operator onto the nearest stable Koopman operator. In this context, they describe how the eigenvalues of the Koopman operator can be used to assess the stability of the underlying system. However, this statement is formulated under the assumption of a known and exact Koopman operator. Furthermore, in the context of observer design with EDMD, it has already been derived how the observability of a nonlinear system can be determined by a Koopman operator-based observer form under the strict assumption that the state vector can be formulated as a linear combination of the assumed to be known Koopman eigenfunctions \cite{SB16}. In the vast majority of complex control systems, these prerequisites cannot be considered to have been met, neither a known exact Koopman operator nor known Koopman eigenfunctions. Therefore, in this work we focus on evaluating the mentioned theoretical results using numerically approximated systems that do not satisfy these conditions, assuming them to be accurate enough for this purpose. 

%The authors in \cite{YLH17} provide a method for nonlinear input-output model reduction using Koopman operators to construct controllability and observability gramians for EDMD approximated systems.
%The authors in \cite{GP17} refer to dynamical systems with the property of bilinearity, which do not have a major impact in control engineering applications.
The aim of this work is threefold: (1) Review of the Koopman operator theory and Extended Dynamic Mode Decomposition, (2) Study of academic examples in a model-based manner and results on a mechatronic test bench with regard to prediction accuracy, stability, controllability and observability, (3) Discussion of the applicability of Koopman operator-based data-driven methods to practical control engineering systems and their subsequent conclusion. 

\section{Background}\label{sec:background}
In the following, we introduce  (\ref{subsec:theory}) the basics of Koopman operator theory and (\ref{subsec:edmd}) Extended Dynamic Mode Decomposition (EDMD). Moreover, it is described (\ref{subsec:edmdc}) how to extend the EDMD method to systems with control and (\ref{subsec:prior_knowledge}) how to take prior knowledge into account. In the following, the index $t$ means in all cases that we are dealing with discrete-time system descriptions.
\subsection{Koopman Operator Theory}\label{subsec:theory}
The following basic theory is taken from \cite{BK19}. Consider a continuous-time autonomous dynamical system
\begin{equation}
	\dfrac{d}{dt} \vec{x}=\vec{f}(\vec{x}), 
\end{equation}
where $\vec{x}\in\mathbb{R}^n$ is the state and $\vec{f}$ is a Lipschitz continuous function. For a given time $t$, we may consider the flow map $\vec{F}_t:\mathbb{R}^n\rightarrow\mathbb{R}^n$, which maps the state $\vec{x}(t_0)$ forward in time to $\vec{x}(t_0+t)$, according to
\begin{equation}
	\vec{x}(t_0+t)=\vec{F}_t(\vec{x}(t_0))=\vec{x}(t_0)+\int_{t_0}^{t_0+t}\vec{f}(\vec{x}(\tau))d\tau.
\end{equation}
In particular, this induces a discrete-time dynamical system 
\begin{equation}
	\vec{x}_{k+1}=\vec{F}_t(\vec{x}_k),
\end{equation}
where $\vec{x}_k=\vec{x}(k\Delta t)$.

The Koopman operator advances measurement functions of the state with the flow of the dynamics. We consider \textit{observables}, which are functions $g:\mathbb{R}^n\rightarrow\mathbb{R}$ that can be thought of as the output variables of a system.  
The Koopman operator $\mathcal{K}_t$ is an infinite-dimensional linear operator that acts on observable functions $g$ as
\begin{equation}
	\mathcal{K}_t g=g\circ \vec{F}_t,
\end{equation}
where $\circ$ is the composition operator. For a discrete-time system with step size $\Delta t$, this becomes
\begin{equation}
	\mathcal{K}_t g(\vec{x}_k )=g(\vec{F}_t (\vec{x}_k ))=g(\vec{x}_{k+1}).
\end{equation}
In other words, the Koopman operator defines an infinite-dimensional linear dynamical system that advances the observation of the state $g(\vec{x}_k)$ to the next time step. This is true for any observable function $g$ and any state $\vec{x}_k$.

For sufficiently smooth dynamical systems, it is also possible to define the continuous-time analogue of the Koopman dynamical system
\begin{equation}
	\dfrac{d}{dt}g=\mathcal{K}g.
\end{equation}
The operator $\mathcal{K}$ is the infinitesimal generator of the one-parameter family of transformations $\mathcal{K}_t$. It is defined by its action on an observable function
\begin{equation}
	\mathcal{K}g=\lim_{t\rightarrow 0}{\dfrac{\mathcal{K}_t g-g}{t}}=\lim_{t\rightarrow 0}{\dfrac{g\circ \vec{F}_t -g}{t}}.
\end{equation}
$\mathcal{K}$ and $\mathcal{K}_t$ are related by $\mathcal{K}=\log(\mathcal{K}_t)/\Delta t$.

A Koopman invariant subspace has the property that all functions in this subspace remain in this subspace after being acted on by the Koopman operator. For the vast majority of systems, it is not easy to find observable functions that directly span a Koopman invariant subspace \cite{BBPK16}.

\subsection{Extended Dynamic Mode Decomposition}\label{subsec:edmd}
For the approximation of the Koopman operator, observable functions $\vec{\Psi}(\vec{x})$ must first be defined
\begin{equation}
	\vec{\Psi}(\vec{x})=\begin{pmatrix}
		\psi_1(\vec{x})&\psi_2(\vec{x})&\cdots&\psi_N(\vec{x})
	\end{pmatrix}^\top,
\end{equation}
where $N$ is the number of observable functions.

\begin{figure*}[!t]
	\centering
	\subfloat[straight EDMD prediction \label{fig:edmd}]{\includegraphics{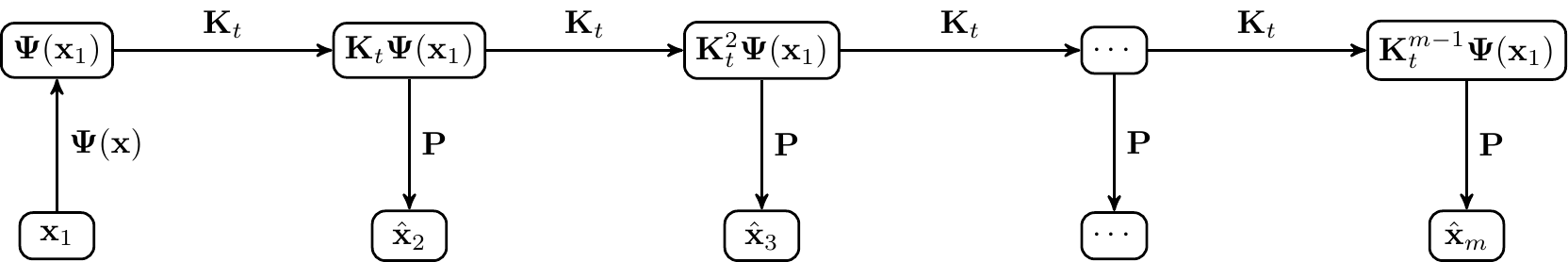}}
	\\
	\centering
	\subfloat[EDMD prediction with correction \label{fig:edmd_corr}]{\includegraphics{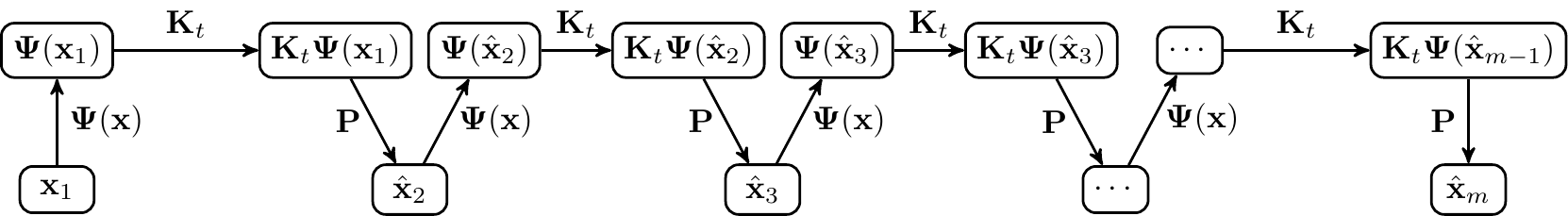}}
	\caption{Resulting system description for (a) straight EDMD prediction and (b) EDMD prediction with correction.}
\end{figure*}

The measurement data, which includes $M$ measurement points and is used to train the EDMD model, needs to be arranged into snapshot matrices
\begin{align}
	\label{snapshotx}
	\vec{X}&=\begin{pmatrix}
		\vec{x}_1& \vec{x}_2&\cdots&\vec{x}_{M-1}
	\end{pmatrix}&\in \mathbb{R}^{n\times (M-1)},\\
	\vec{X}'&=\begin{pmatrix}
		\vec{x}_2& \vec{x}_3&\cdots&\vec{x}_M
	\end{pmatrix}&\in \mathbb{R}^{n\times (M-1)}.
\end{align}
The measurement data do not necessarily have to originate from a single measurement or simulation; pairs from two successive snapshots entered in the correct columns in $\vec{X}$ and $\vec{X}'$ are sufficient.\\
Evaluating the observable functions for the snapshot matrices yields
\begin{align}
	\label{snapshotpsi}
	\vec{\Psi}(\vec{X})&=\begin{pmatrix}
		\vec{\Psi}(\vec{x}_1)&\cdots&\vec{\Psi}(\vec{x}_{M-1})	
	\end{pmatrix}\in\mathbb{R}^{N\times (M-1)},\\
	\vec{\Psi}(\vec{X'})&=\begin{pmatrix}
		\vec{\Psi}(\vec{x}_2)&\cdots&\vec{\Psi}(\vec{x}_M)	
	\end{pmatrix}\in\mathbb{R}^{N\times (M-1)}.
\end{align}
The approximated Koopman operator $\vec{K}_t\in\mathbb{R}^{N\times N}$ corresponds to the linear dynamics of the observable functions, yielding
\begin{equation}
	\vec{\Psi}(\vec{X'})\approx\vec{K}_t\vec{\Psi}(\vec{X})\Rightarrow \vec{K}_t=\vec{\Psi}(\vec{X'})\vec{\Psi}^+(\vec{X}),
\end{equation}
where $\vec{\Psi}^+$ denotes the Moore-Penrose inverse of matrix $\vec{\Psi}$. It can be shown that $\vec{K}_t\rightarrow\mathcal{K}_t$, if $M\rightarrow\infty$ \cite{KKS16} and $N\rightarrow\infty$ \cite{KM18c}.\\
The resulting system description for a straight EDMD prediction step is given by
\begin{equation}\label{eq:edmd_pred}
	\hat{\vec{\Psi}}(\vec{x}_{k+1})=\vec{K}_t\vec{\Psi}(\vec{x}_k)
\end{equation}
and shown in Fig. \ref{fig:edmd}. Here, the hat on the symbols means in each case that it is an estimated or predicted quantity because in general, for systems that do not span a Koopman invariant subspace, it is
\begin{equation}
	\vec{\Psi}(\vec{x}_{k+1})\neq\vec{K}_t\vec{\Psi}(\vec{x}_k).
\end{equation}
For the extraction of $\vec{x}$, a projection matrix $\vec{P}$ is needed. Therefore it is advantageous if the first $n$ functions in $\vec{\Psi}(\vec{x})$ contain the original state vector
\begin{equation}
	\vec{\Psi}(\vec{x})=\begin{pmatrix}
		\vec{x}&\psi_{n+1}(\vec{x})&\cdots&\psi_N(\vec{x})
	\end{pmatrix}^\top,
\end{equation} 
yielding
\begin{equation}
	\vec{x}_{k+1}=\vec{P}\vec{\Psi}(\vec{x}_{k+1}) \text{ with } \vec{P}=\begin{pmatrix}
		\vec{I}_n& \vec{0}_{n\times (N-n)}
	\end{pmatrix},
\end{equation}
where $\vec{I}_n$ is the $n\times n$ identity matrix and $\vec{0}_{n\times(N-n)}$ is the $n\times(N-n)$ zero matrix.

In most cases, a Koopman invariant subspace is not spanned by $\vec{\Psi}(\vec{x})$. Therefore, a prediction by \eqref{eq:edmd_pred} will be subject to error for a longer time horizon. An alternative for a more accurate prediction is to apply a correction to the states of $\vec{\Psi}(\vec{x})$ in each computation step, as described in \cite{KM18b} and here referred to as EDMD prediction with correction. This extracts the state vector $\vec{x}$ at each step and thus reevaluates the observable function, so it no longer matters that $\vec{\Psi}(\vec{x})$ does not span a Koopman invariant subspace:
\begin{equation}\label{eq:edmd_corr_pred}
	\hat{\vec{x}}_{k+1}=\vec{P}\hat{\vec{\Psi}}(\vec{x}_{k+1})=\vec{P}\vec{K}_t\vec{\Psi}(\vec{x}_k),
\end{equation}
which is shown in Fig. \ref{fig:edmd_corr}. Note that the EDMD prediction with correction does not actually yield a linear system description because a nonlinear function is evaluated at each calculation step. 
\subsection{Extension of EDMD to Actuated Systems}\label{subsec:edmdc}
The Koopman operator theory and the original EDMD algorithm were originally formulated for autonomous systems. For control systems the EDMD algorithm must be extended by a control input. A common approach to this is to add the system inputs to the observable functions, so that $\vec{\Psi}(\vec{x},\vec{u})$.

Here it is important to eliminate the input dependent components from the observables and lift the dynamics by using the remaining input-free observables, which is explained in \cite{SA21}. Therefore, control-affine systems
\begin{align}
	\dot{\vec{x}}&=\vec{f}(\vec{x})+\vec{B}\vec{u},\\
	\vec{y}&=\vec{g}(\vec{x},\vec{u})
\end{align}
are considered in the following.
The vector $\vec{x}\in\mathbb{R}^n$ denotes the states of the system and $\dot{\vec{x}}\in\mathbb{R}^n$ the time derivative of the states. $\vec{u}\in\mathbb{R}^p$ denotes the system inputs, $\vec{B}\in\mathbb{R}^{n\times p}$ the constant input matrix and $\vec{y}\in\mathbb{R}^q$ the system outputs. The restriction to input-affine systems is legitimate in the control engineering context because most engineering systems exhibit this property. 

The EDMD system with control can then be described by stacking $\vec{\Psi(x)}$ and $\vec{u}$, so that the dynamics become
\begin{equation}
	\vec{\Psi}(\vec{x}_{k+1})\approx\vec{K}_t\vec{\Psi}(\vec{x}_k)+\vec{B}_t\vec{u}_k=\begin{pmatrix}
		\vec{K}_t&\vec{B}_t
	\end{pmatrix}\begin{pmatrix}
		\vec{\Psi(\vec{x}_k)}\\ \vec{u}_k
	\end{pmatrix},
\end{equation}
which is similar to the approach of DMDc (Dynamic Mode Decomposition with Control) \cite{PBK16}. 
With data results
\begin{align}
	\vec{\Psi(X')}&\approx\vec{K}_t\vec{\Psi(X)}+\vec{B}_t\vec{U}
	=\begin{pmatrix}
		\vec{K}_t&\vec{B}_t
	\end{pmatrix}\begin{pmatrix}
		\vec{\Psi(X)}\\ \vec{U}
	\end{pmatrix},
\end{align}
thus
\begin{equation}
	\begin{pmatrix}
		\vec{K}_t&\vec{B}_t
	\end{pmatrix}=\vec{\Psi(X')}\begin{pmatrix}
		\vec{\Psi(X)}\\\vec{U}
	\end{pmatrix}^+,
\end{equation}
where
\begin{equation}
	\vec{U}=\begin{pmatrix}
		\vec{u}_1&\vec{u}_2\cdots \vec{u}_{M-1}
	\end{pmatrix}\in\mathbb{R}^{p\times(M-1)}
\end{equation}
is the measurement of the system input signal and $\vec{B}_t\in\mathbb{R}^{N\times p}$. 
The resulting system description for a straight EDMD prediction step is given by
\begin{equation}\label{eq:edmdc_pred}
	\vec{\hat{\Psi}}(\vec{x}_{k+1})=\vec{K}_t\vec{\Psi}(\vec{x}_k)+\vec{B}_t\vec{u}_k,	
\end{equation}
respectively the EDMD prediction with correction is given by
\begin{equation}
	\vec{\hat{x}}_{k+1}=\vec{P}\vec{\hat{\Psi}}(\vec{x}_{k+1})=\vec{P}\left(\vec{K}_t\vec{\Psi}(\vec{x}_k)+\vec{B}_t\vec{u}_k\right).
\end{equation}

\subsection{Taking Prior Knowledge Into Account}\label{subsec:prior_knowledge}
The choice of the observable functions is crucial for the success of the approximation result in the EDMD method. With the motivation to combine the physically motivated models with the data-driven models in a sensible way, it is advantageous to take prior knowledge into account. The authors in \cite{MCTM19} motivate the use of time derivatives of the system dynamics, if known, in the observable functions. The justification for this is based on the connection that the observables must contain all higher dimensional derivatives of the original states in order to span a Koopman invariant subspace \cite{BBPK16}. This approach is formulated using a Taylor series approximation and thus error bounds can be derived \cite{MCTM20}. The authors in \cite{NSKZ21} adopted this idea and formalized it as \textit{EDMD-Lie} being applicable for a few nonlinear elementary functions. In \cite{CHM+20}, this strategy was transferred to the case where no prior knowledge exists, by estimating the higher derivatives of the states using high gain observers. In the following, we assume that basic prior knowledge of the systems, e.g., the form of nonlinearities, is known and choose the observable functions accordingly. 

\section{Analysis of Academic Examples}\label{sec:academic_examples}
To illustrate, we first describe the simulative application of the EDMD algorithm to nonlinear academic examples. We cover the nonlinear pendulum and the Duffing oscillator as these are well known nonlinear systems with several isolated fixed points. For both examples, we first compute a data-driven EDMD model. We then analyze the prediction accuracy depending on the observable functions and the training data. In addition, we address how system properties such as stability, controllability, and observability are reflected in the EDMD model.

\subsection{Nonlinear Pendulum}\label{subsec:pendulum}
The nonlinear pendulum can be described by the following differential equations
\begin{align}
	\dot{x}_1&=x_2,\\
	\dot{x}_2&=-\sin(x_1)-dx_2+u,\\
	y&=x_1,
\end{align}
where $x_1$ and $x_2$ are the angle and angular velocity of the pendulum, respectively, and a low damping $d=0.05$ is assumed. The system input $u$ is the torque on the pendulum. 

For the generation of the training data, 100 simulated (numerical integration with RK4) trajectories with the duration of \SI{3}{\second} each with a step size of $\Delta t=\SI{0.01}{\second}$ and random initial conditions from the basin of attraction of the stable equilibrium $\vec{x}^*=\begin{pmatrix}
	0&0
\end{pmatrix}^\top$ and first with $u=0$ were used.\\
According to Sec. \ref{subsec:prior_knowledge}, the observable functions
\begin{equation}
	\vec{\Psi}(\vec{x})=\begin{pmatrix}
		x_1&x_2&\sin(x_1)&\cos(x_1)x_2&\cdots
	\end{pmatrix}^\top
\end{equation}		
are used. 
\begin{figure}[!t]
	\centering
	\subfloat[$N=6$]{\includegraphics{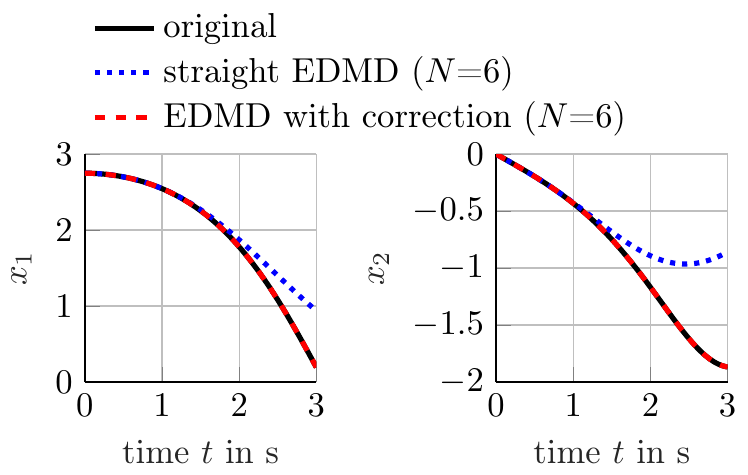}}
	\\
	\subfloat[$N=24$]{\includegraphics{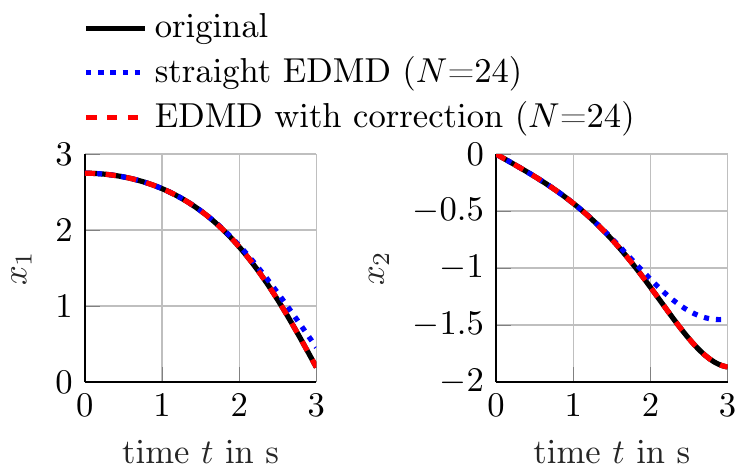}}
	\caption{Straight EDMD prediction and EDMD prediction with correction of the nonlinear pendulum using (a)~6 or (b)~24 observable functions.}
	\label{fig:exTrajPendulum}
\end{figure}

To evaluate the resulting prediction accuracy, an exemplary initial condition was investigated using both straight EDMD prediction and EDMD prediction with correction and compared to the numerical simulation of the original nonlinear system, as shown in Fig. \ref{fig:exTrajPendulum}. In addition, the number of observable functions was varied here, respectively. Note that the initial condition $\vec{x}(0)=\begin{pmatrix}
	7\pi/8&0
\end{pmatrix}^\top$ is outside the small angle range. It can be seen that the EDMD method with correction term has a very high prediction accuracy.  However, the straight EDMD method approximates the trajectory of the original system very well only locally, i.e. for a short time horizon, and then deviates. This can be explained by the fact that the chosen library does not span a Koopman invariant subspace and thus not all states in the observable functions can be predicted correctly based on them. Increasing the number of observables causes the prediction to deviate later in time, see Fig. \ref{fig:exTrajPendulum}~(b).

\begin{figure}[!t]
	\centering
	\includegraphics{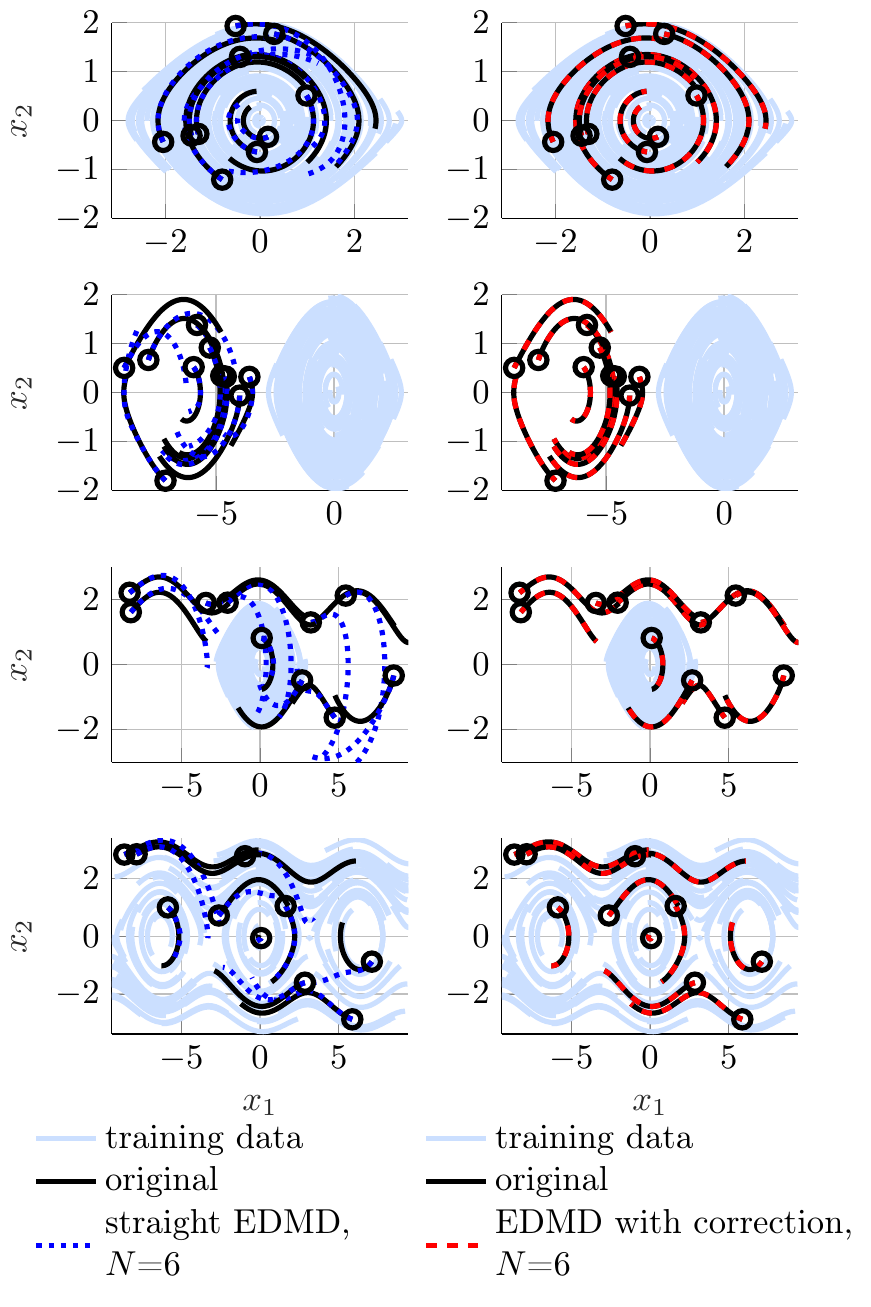}
	\caption{Impact of training data selection on prediction accuracy for the pendulum. The initial conditions are marked by black circles.}
	\label{fig:phasePlotPendulum}
\end{figure}

In addition, selecting the amount of training data has an impact on the prediction accuracy of the test data. Fig. \ref{fig:phasePlotPendulum} shows various combinations of training data (100 simulated trajectories with the duration of \SI{3}{\second} each with a step size of $\Delta t=\SI{0.01}{\second}$) and test data (10 example test trajectories) of the pendulum. The following observations can be made:
\begin{itemize}
	\item The EDMD prediction method with correction term provides extremely high prediction accuracy while being very robust to the selection of training data and the basins of attraction of different equilibrium positions.
	\item The straight EDMD prediction method provides locally high prediction accuracy, also with respect to the mapping of different basins of attraction; however, the predicted trajectory deviates from that of the original system after a short time, as already described above.
	\item The prediction accuracy of the straight EDMD prediction method depends sensitively on the selection of the training data. It follows that the training data should cover as large a range of the state space as possible to increase the prediction accuracy.
\end{itemize}
The considered system of the nonlinear damped pendulum is asymptotically stable in the basin of attraction. This property is reflected in the continuous-time eigenvalues
\begin{align}
	\lambda_{1,2}&=-0.059\pm i2.006,\\
	\lambda_{3,4}&=-0.009\pm i0.759,\\
	\lambda_5&=-0.085,\\
	\lambda_6&=-0.052
\end{align}
of the approximated EDMD system, which all have negative real parts. Thus, a strong connection between the original nonlinear system and the EDMD approximated system can be seen here. Note that the nonlinear system of the pendulum actually exhibits a continuous eigenvalue spectrum \cite{LKB18}, while EDMD is limited to map only discrete eigenvalues.

Next, the properties of controllability and observability of the EDMD model are examined. The  nonlinear system is controllable and observable, which can be seen very easily \cite{Kha15}. The EDMD model was trained as described in \ref{subsec:edmdc} where the training data was generated in the same way as above and additional random input signals ${u\in\left[-1,1\right]}$ were added. This results in the following continuous-time controllability matrix of the EDMD system
\begin{equation}
	\mathrm{rank}\mathcal{C}_N=\mathrm{rank}\begin{pmatrix}
		\vec{B}&\vec{K}\vec{B}&\cdots&\vec{K}^{N-1}\vec{B}
	\end{pmatrix}=N
\end{equation}
and the continuous-time observability matrix of the EDMD system
\begin{equation}
	\mathrm{rank}\mathcal{O}_N=\mathrm{rank}\begin{pmatrix}
		\vec{C}\\ \vec{C}\vec{K}\\\vdots\\ \vec{C}\vec{K}^{N-1}
	\end{pmatrix}=N,
\end{equation}
which both have full rank for different numbers ${N=2,\cdots,24}$ of observable functions. Therefore, a strong connection between the original nonlinear system and the approximated EDMD model can be seen.

\subsection{Duffing Oscillator}
The dynamics of the Duffing oscillator can be described by the following differential equations
\begin{align}
	\dot{x}_1&=x_2,\\
	\dot{x}_2&=x_1-x_1^3-\delta x_2+u,\\
	y&=x_1,
\end{align}
where $x_1$ and $x_2$ are the displacement and velocity of a mass, respectively, and a low damping $\delta=0.1$ is assumed. The system input $u$ is an external force acting.

For the generation of the training data, 100 simulated (numerical integration with RK4) trajectories with the duration of \SI{3}{\second} each with a step size of $\Delta t=\SI{0.01}{\second}$ and random initial conditions in the region $-2\leq x_1,x_2\leq2$ beginning with $u=0$ were used.\\
According to Sec. \ref{subsec:prior_knowledge}, the observable functions
\begin{equation}
	\vec{\Psi}(\vec{x})=\begin{pmatrix}
		x_1&x_2&x_1^3&x_1^2 x_2&x_1^5&x_1 x_2^2&\cdots
	\end{pmatrix}^\top
\end{equation}		
are used. 

\begin{figure}[!t]
	\centering
	\subfloat[$N=6$]{\includegraphics{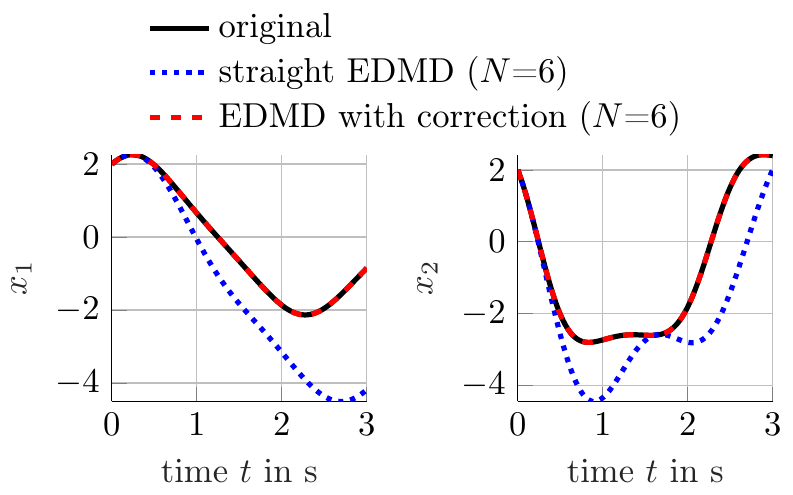}}
	\label{subfig:exTrajDuffing_K_6}
	\subfloat[$N=20$]{\includegraphics{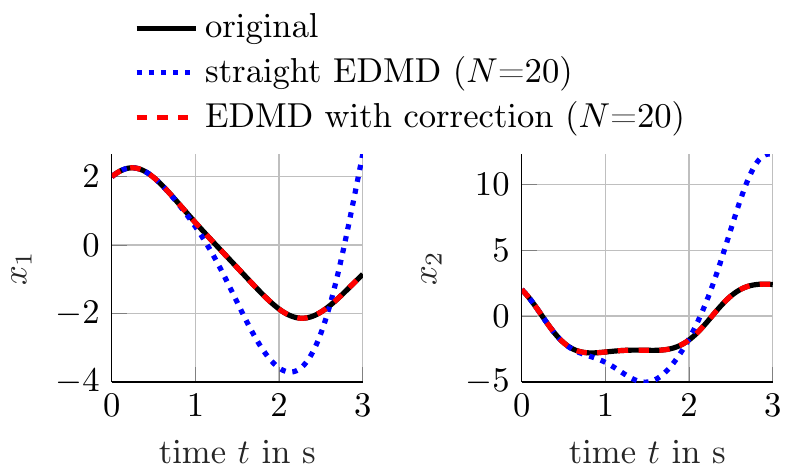}}
	\label{subfig:exTrajDuffing_K_20}
	\caption{Straight EDMD prediction and EDMD prediction with correction of the Duffing oscillator using (a)~6~or~(b)~20 observable functions.}
	\label{fig:exTrajDuffing}
\end{figure}

\begin{figure}[!t]
	\centering
	\includegraphics{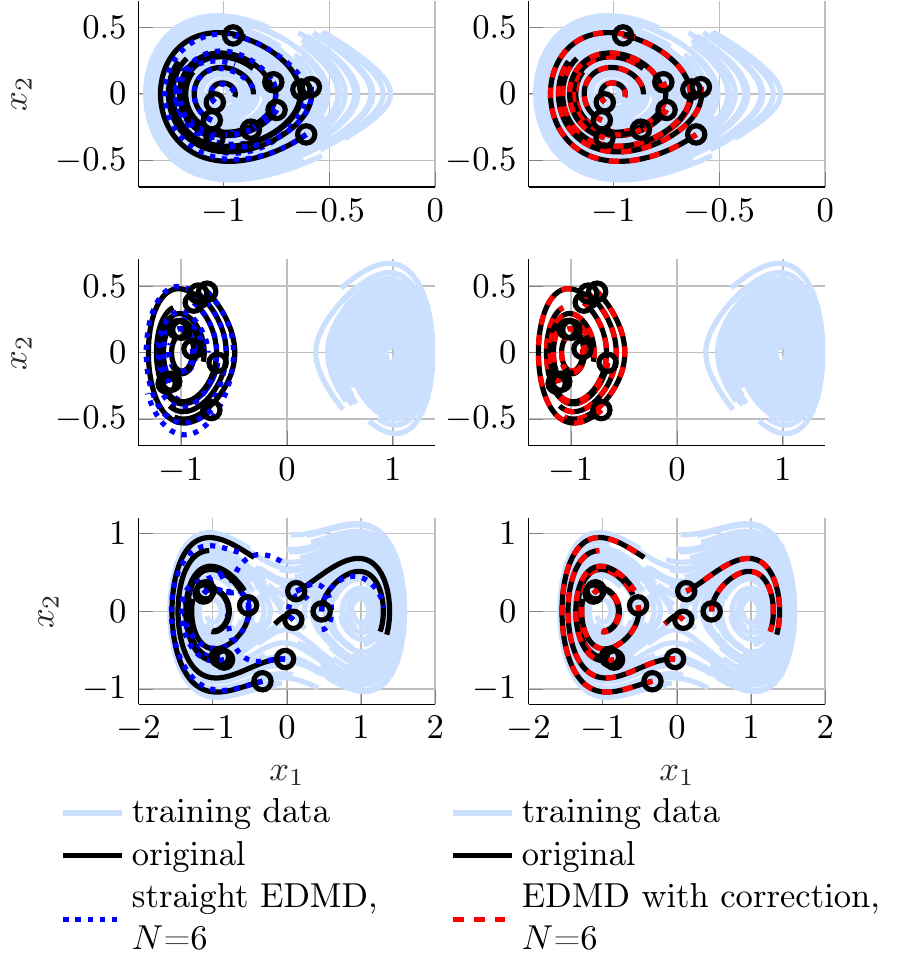}
	\caption{Impact of training data selection on prediction accuracy for the Duffing oscillator. The initial conditions are marked by black circles.}
	\label{fig:phasePlotDuffing}
\end{figure}

With the Duffing oscillator, the same observations, see Fig.\,\ref{fig:exTrajDuffing} and Fig.\,\ref{fig:phasePlotDuffing}, can be made as with the pendulum.

The continuous-time eigenvalues have the following values
\begin{align}
	\lambda_{1,2}&=-0.189\pm i4.009,\\
	\lambda_{3,4}&=-0.120\pm i1.172,\\
	\lambda_5&=-0.080,\\
	\lambda_6&=-0.116,
\end{align}
where the negative real parts imply stability as well as the nonlinear system. 

With additional random input signals ${u\in\left[-1,1\right]}$, the criteria of controllability and observability, which are met in the original nonlinear system \cite{Kha15}, can also be found in the EDMD model 
\begin{equation}
	\mathrm{rank}\mathcal{C}_N=\mathrm{rank}\mathcal{O}_N=N
\end{equation}
for $N=2,\cdots,20$. 
\section{Experimental Results}\label{sec:test_bench}
\begin{figure}[!t]
	\centering
	\includegraphics[width=4cm]{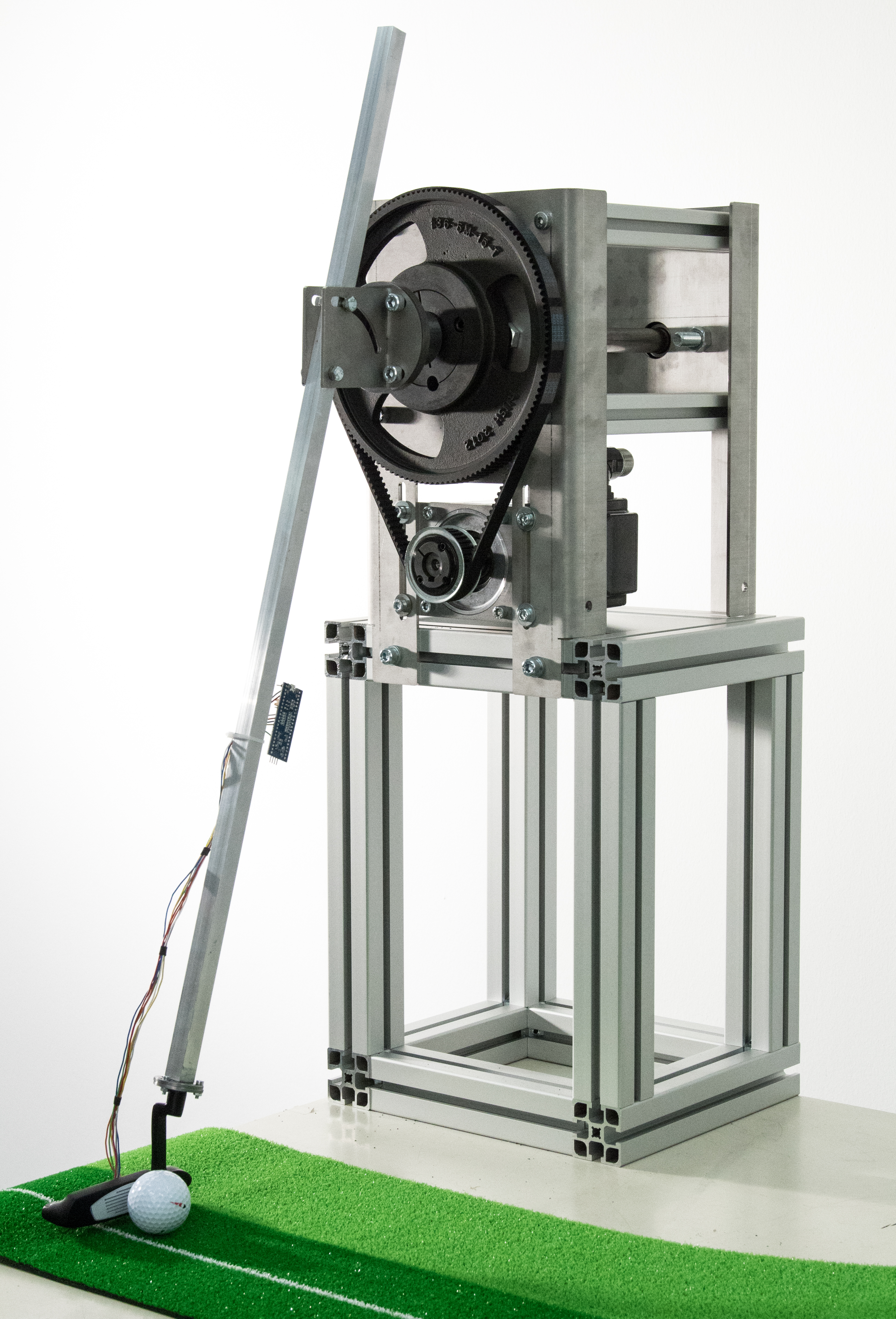}
	\caption{Golf robot as a demonstrator for data-driven methods in control engineering.}
	\label{fig:golfrobot}
\end{figure}

The golf robot shown in Fig. \ref{fig:golfrobot} is being developed as a demonstrator for data-driven methods in control engineering, where the goal is for the robot to independently learn to putt the ball from a point on the green after a few test strokes on an unknown grass surface.
The stroke mechanism consists of two gear shafts connected with a toothed belt drive. The drive is located on the lower gear shaft and the golf club is mounted on the upper gear shaft. A simplified nonlinear model combines the masses into a single rigid body with torque $u$ as control input.  The differential equations with parameters shown in Table \ref{tab:parameters} can be described by the following:
\begin{align}
	\dot{x}_1&=x_2,\\
	\dot{x}_2&=\dfrac{-mga\sin{x_1}-M_d(\vec{x})+4u}{J}
\end{align} 
where $\vec{x}=\begin{pmatrix}
	\varphi&\dot{\varphi}
\end{pmatrix}^T$ contains the angle and angular velocity of the golf club and the nonlinear damping torque 
\begin{equation}
	M_d(\vec{x})=dx_2+r\mu\mathrm{sgn}{x_2}\lvert mx_2^2 a+mg\cos{x_1}\rvert
\end{equation}
is motivated by a combination of Coulomb friction and sliding friction. 
\begin{table*}[!htbp]
	\caption{Physical parameters of the golf robot.}
	\label{tab:parameters}
	\begin{center}
		\begin{tabular}{cll}
			\hline
			symbol&physical parameter&value\\
			\hline
			$m$& mass of the golf club&  \SI{0.5241}{\kilo\gram}\\ 
			$J$& inertia of the golf club&  \SI{0.1445}{\kilo\gram\per\meter^2}\\
			$g$& gravity constant& \SI{9.81}{\meter\per\second^2}\\
			$a$& length from the axis of rotation to the center of mass of the golf club& \SI{0.4702}{\meter}\\
			$d$& dynamic friction constant& \SI{0.0132}{\kilo\gram\meter^2\per\second}\\
			$r$& length from the axis of rotation to the friction point& \SI{0.0245}{\meter}\\
			$\mu$& static friction constant& $1.5136$\\
			\hline
		\end{tabular}
	\end{center}
\end{table*}
This results in the following observable functions
\begin{equation}
	\vec{\Psi}(\vec{x})=\begin{pmatrix}
		x_1&x_2&\sin x_1&\mathrm{sgn}{x_2}\lvert mx_2^2a+mg\cos{x_1}\rvert
	\end{pmatrix}^\top.
\end{equation}
According to the above conclusions for the training data, several measurements with different input signals
\begin{itemize}
	\item chirp signal of $u$
	\item sine signal of $u$
	\item step signal of $u$
\end{itemize}
and different amplitudes were used and combined in the matrices $\vec{X}$ and $\vec{X'}$ with a $\SI{1}{\kilo\hertz}$ sampling rate. 
The output variable of the system, which is measured directly, is the angle $y=x_1=\varphi$ of the golf club. Training data for $x_2$ were generated offline by smoothing spline interpolation followed by numerical differentiation of $x_1$.   

\begin{figure*}[!t]
	\centering
	\includegraphics{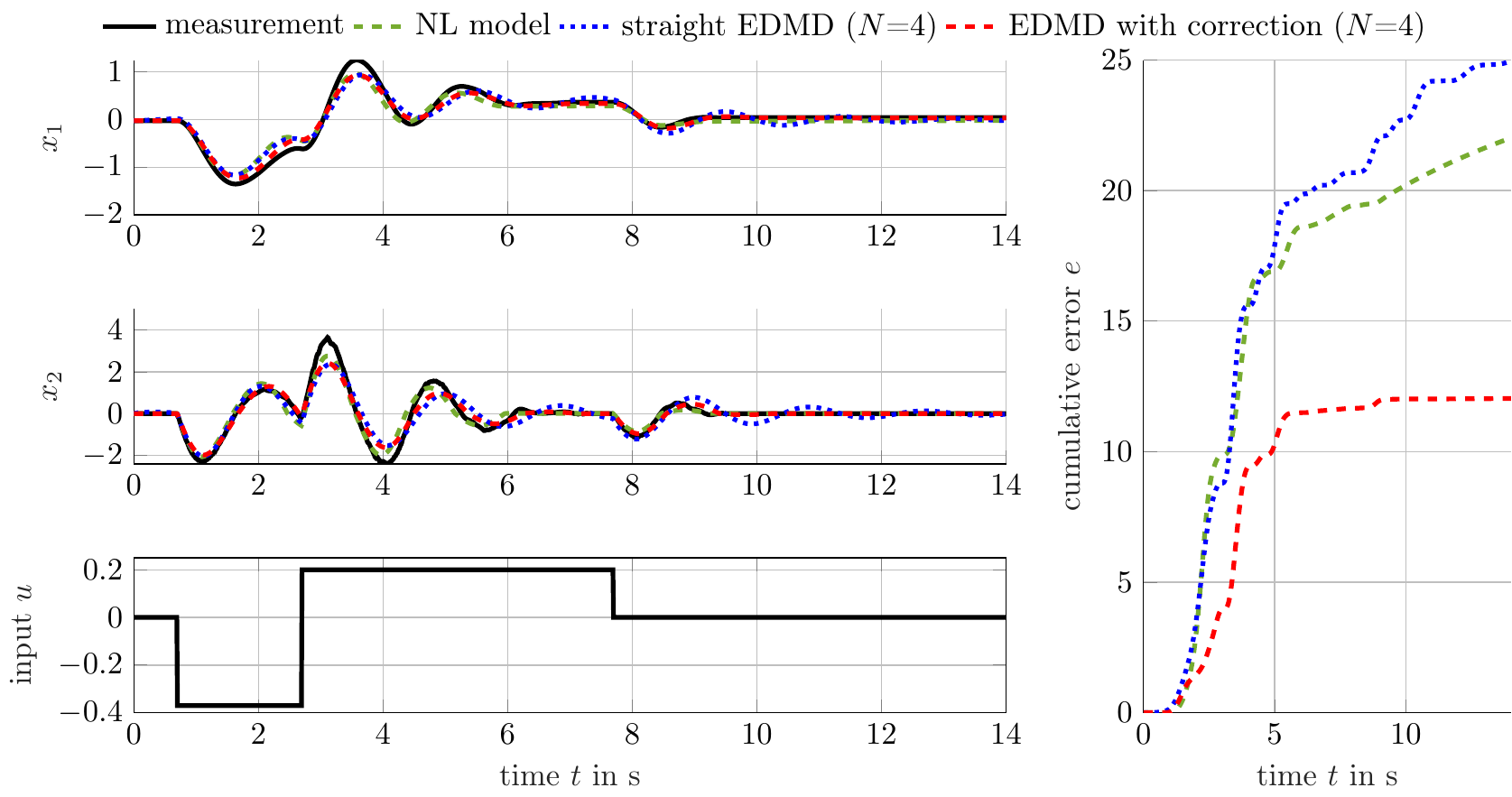}
	\caption{Prediction accuracy of EDMD models in comparison with the nonlinear parameterized model of the golf robot.}
	\label{fig:golfrobot_EDMD}
\end{figure*}

After that, an unseen trajectory was used as a test trajectory to evaluate the prediction accuracy of the EDMD models, see Fig. \ref{fig:golfrobot_EDMD}. The prediction over time is shown on the left and the cumulative prediction error
\begin{equation}
	e(t_k)=\sum_{m=1}^{k} (x_{1,\text{meas}}(t_m)-x_{1,\text{pred}}(t_m))^2
\end{equation}
of $y=x_1$ is shown on the right. 

In green, the numerical integration of the nonlinear parameterized model is also shown. It can be seen that both EDMD models provide good prediction accuracy. Apparently, the EDMD prediction with correction is able to map the nonlinear friction from $t=\SI{8}{\second}$ a little more precisely than the straight EDMD prediction. 

Similar to the pendulum in Sec. \ref{subsec:pendulum}, the system is stable in the basin of attraction of the stable equilibrium. This property is reflected in the continuous-time eigenvalues 
\begin{align}
	\lambda_1&=-13.912,\\
	\lambda_2&=-0.146,\\
	\lambda_{3,4}&=-0.498\pm i3.395
\end{align}
of the EDMD model, which all have negative real parts.

The system is controllable and observable and the matrices of controllability and observability for the EDMD model
\begin{equation}
	\mathrm{rank}\mathcal{C}_N=\mathrm{rank}\mathcal{O}_N=N
\end{equation}
for $N=4$ have full rank. This means that the results of the analysis of the academic examples could also be reproduced using measurements from a mechatronic test bench. 

\section{Discussion of Applicability for Control Engineering}\label{sec:discussion}
EDMD provides a way to accurately approximate a nonlinear dynamical system as a linear model in a data-driven manner assuming basic prior knowledge. This is promising for applications in control engineering as it eliminates the need for detailed physical modeling with subsequent parameter identification and model adaptation, which is a major effort in the control design process. 

The prediction accuracy of the EDMD model depends on the one hand on the selected observable functions and the amount of training data. On the other hand, the prediction method plays an important role. In the straight EDMD method, a linear system in state-space representation is present, which allows a prediction simply by a matrix-vector-product operation, but provides high accuracy only for a small time horizon and for wisely selected training date. In contrast, EDMD with correction term provides a high accuracy and is even robust to the amount of training data. However, the computational effort is higher with this method, because a second matrix-vector product and a nonlinear function evaluation are additionally required in each calculation step, so that no actual linear description results. When used in a model predictive control, the selection of the prediction method should therefore depend on how fast the prediction must be executed and which level of accuracy is demanded. 

For the practical application in control engineering, it is crucial that the underlying process models correctly represent relevant system properties. Based on the applications in this work, it was demonstrated simulatively as well as experimentally that stability as well as controllability and observability are reflected in the EDMD models. Given prior knowledge of the nonlinear system, these properties can be easily verified.

Using the example of the golf robot as a mechatronic test bench, it was demonstrated that the EDMD model reproduces the nonlinear system dynamics comparably well as the nonlinear parameterized model and correctly represents relevant system-theoretical properties.

\section{Conclusion \& Outlook}\label{sec:conclusion}
In summary, this work has established the applicability of EDMD models on practical control engineering applications. The investigated EDMD models enable high prediction accuracy and correctly represent critical system theory properties such as stability, controllability, observability, as determined simulatively on academic examples and experimentally on a mechatronic test bench. With the help of the EDMD method, a linear system description of a nonlinear system results directly from measurement data and prior knowledge, so that the effort for model building and fitting is significantly reduced compared to the classical approach, which offers enormous potential for practical control engineering applications.

To make the model more physically plausible, the future research is how to integrate principles of energy conservation into the modeling. The next step is then to investigate how the learned EDMD model can be used for subsequent controller design, see Fig. \ref{fig:overview}. For this purpose, the computational effort in prediction should be investigated in more detail and it should be analyzed whether the effect of the linear system description is ruined by a too high number of observable functions.
\section*{Acknowledgment}
This work was developed in the junior research group DART (Daten\-ge\-trie\-be\-ne Methoden in der Regelungstechnik), Paderborn University, and funded by the Federal Ministry of Education and Research of Germany (BMBF - Bundes\-ministerium für Bildung und Forschung) under the funding code 01IS20052. The responsibility for the content of this publication lies with the authors.

%\bibliographystyle{IEEEtran}
%\bibliography{IEEEabrv,bibliography}

% Generated by IEEEtran.bst, version: 1.14 (2015/08/26)

\end{document}